\documentclass[showpacs,twocolumn,floatfix,superscriptaddress]{revtex4}
\usepackage{epsfig}
\usepackage{graphicx}
\usepackage{amsmath}
\begin{document}
\title{Scattering theory of plasmon-assisted entanglement transfer and distillation}
\author{J.L. van Velsen}
\affiliation{Instituut-Lorentz, Universiteit Leiden, P.O. Box 9506, 2300 RA Leiden, The Netherlands}
\author{J. Tworzyd{\l}o}
\affiliation{Instituut-Lorentz, Universiteit Leiden, P.O. Box 9506, 2300 RA Leiden, The Netherlands}
\affiliation{Institute of Theoretical Physics, Warsaw University, Ho\.{z}a 69, 00-681 Warszawa, Poland}
\author{C.W.J. Beenakker}
\affiliation{Instituut-Lorentz, Universiteit Leiden, P.O. Box 9506, 2300 RA Leiden, The Netherlands}
\date{November 2002}
\begin{abstract}
We analyse the quantum mechanical limits to the plasmon-assisted entanglement transfer observed by E. Altewischer, M.P. van Exter,
and J.P. Woerdman [Nature, \textbf{418}, 304 (2002)]. The maximal violation $S$ of Bell's inequality at the photodetectors behind
two linear media (such as the perforated metal films in the experiment) can be described by two ratio's $\tau_{1}$, $\tau_{2}$
of polarization-dependent transmission probabilities. A fully entangled incident state is transferred without degradation for
$\tau_{1}=\tau_{2}$, but a relatively large mismatch of $\tau_{1}$ and $\tau_{2}$ can be tolerated with a small reduction of $S$. We
predict that fully entangled Bell pairs can be distilled out of partially entangled radiation if $\tau_{1}$ and $\tau_{2}$ satisfy a
pair of inequalities.
\end{abstract}
\maketitle

The motivation for this work came from the recent remarkable demonstration by Altewischer, Van Exter, and Woerdman
of the transfer of quantum mechanical entanglement from photons to surface plasmons and back to photons~\cite{Alt02}. 
Since entanglement is a highly fragile property of
a two-photon state, it came as a surprise that this property could survive with little degradation the 
conversion to and from the macroscopic degrees of freedom in a metal~\cite{Bar02}.

We present a 
quantitative description of the finding of Ref.\ ~\cite{Alt02} that the entanglement is lost if it is measured 
during transfer, 
that is to say, if the medium through which the pair of polarization-entangled photons is passed acts as a ``which-way'' detector
for polarization. Our analysis explains why a few percent degradation of entanglement could be realized without requiring a highly
symmetric medium. We predict that the experimental setup of Ref.\ ~\cite{Alt02} could be used to ``distill''~\cite{Ben96,Nie00} fully 
entangled Bell
pairs out of partially entangled incident radiation, and we identify the region in parameter space where this distillation is possible.

We assume that the medium is {\em linear}, so that its effect on the radiation can be described by a scattering matrix. 
The assumption of linearity of the interaction of radiation with surface plasmons is central to the literature on this 
topic~\cite{Sch98, Tre99, Por99,Mar01,Rat88}.
We will not make any specific assumptions on the mode and frequency dependence of the scattering matrix, but extract the smallest
number of independently measurable parameters needed to describe the experiment.
By concentrating on model-independent results we can 
isolate the fundamental quantum mechanical limitations on the entanglement transfer, from the limitations specific for any particular transfer mechanism. 

\begin{figure}[h!]
\epsfig{file=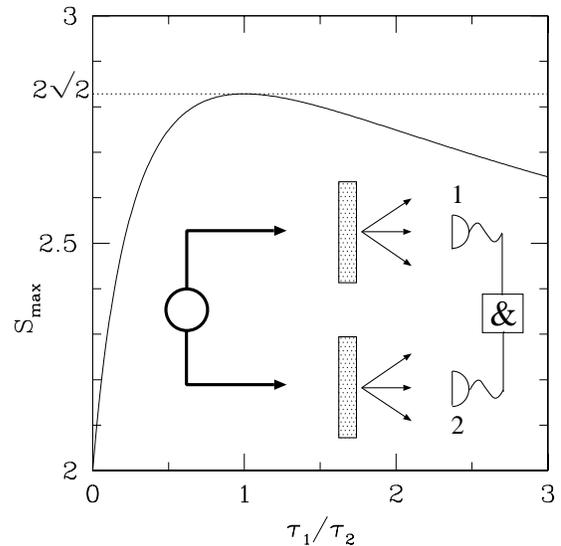,height=8cm,width=8cm}
\caption{Main plot: Efficiency of the entanglement transfer for a fully entangled incident state, as given by 
Eq.\ (\protect\ref{Smaxdef}). The maximal
violation $S_{\rm max}$ of Bell's inequality at the photodetectors is plotted as a function of the ratio $\tau_{1}/\tau_{2}=
T_{1+}T_{2-}/T_{1-}T_{2+}$ of the polarization-dependent transmission probabilities. The inset shows schematically the geometry
of the experiment~\cite{Alt02}. A pair of polarization-entangled photons is incident from the left on two perforated metal films. 
The photodetectors at the right, connected by a coincidence counter, measure the degree of entanglement of the transmitted radiation.} 
\label{smax}
\end{figure}

The system considered is shown schematically in Fig. \ ~\ref{smax}. Polarization-entangled radiation is scattered 
by two objects and detected by a pair of detectors behind the objects in the far-field. The objects used in Ref.\ ~\cite{Alt02} are 
metal films perforated by a square array of subwavelength holes.
The transmission amplitude $t_{\sigma\sigma ',i}$ of object $i=1,2$ relates the transmitted radiation 
(with polarization $\sigma={\rm H},{\rm V}$) to the incident radiation (polarization $\sigma '={\rm H},{\rm V}$). We assume 
a single-mode incident beam and a single-mode detector (smaller than the coherence area), so that we require a set of eight transmission 
amplitudes $t_{\sigma\sigma ',i}$
out of the entire scattering matrix (which also contains reflection amplitudes and transmission amplitudes to other modes).  
The extension to a multi-mode theory (needed to describe some aspects of the experiment~\cite{Alt02}) is left for a future
investigation.

We do not require that the scattering matrix is unitary, so our results remain valid if the objects absorb part of
the incident radiation. What is neglected is the thermal radiation, either from the two objects or from the electromagnetic
environment of the detectors. This thermal noise is insignificant at room temperature and optical frequencies.

The radiation incident on the two objects is in a known, partially entangled state and we wish to determine the degree of entanglement of the 
detected radiation. It is convenient to use a matrix notation. The incident two-photon state has the general form
\begin{equation}
|\Psi_{\rm in}\rangle=a^{\rm in}_{\rm HH}|{\rm HH}\rangle+a^{\rm in}_{\rm HV}|{\rm HV}\rangle+a^{\rm in}_{\rm VH}|{\rm VH}\rangle+a^{\rm in}_{\rm VV}|{\rm VV}\rangle.\label{Psidef}
\end{equation}
The four complex numbers $a^{\rm in}_{\sigma\sigma'}$ form a matrix
\begin{equation}
A_{\rm in}=\left(
\begin{array}{cc}
a^{\rm in}_{\rm HH}&a^{\rm in}_{\rm HV}\\
a^{\rm in}_{\rm VH}&a^{\rm in}_{\rm VV}
\end{array}
\right).\label{Adef}
\end{equation}
Normalization of $|\Psi_{\rm in}\rangle$ requires ${\rm Tr}\,A_{\rm in}^{\vphantom{\dagger}}A_{\rm in}^{\dagger}=1$, with ``Tr'' the trace of a matrix.

The four transmission amplitudes $t_{\sigma\sigma',i}$ of object $i=1,2$ form the matrix
\begin{equation}
T_{i}=\left(
\begin{array}{cc}
t_{{\rm HH},i}&t_{{\rm HV},i}\\
t_{{\rm VH},i}&t_{{\rm VV},i}
\end{array}
\right).\label{Tdef}
\end{equation}
The transmitted two-photon state $|\Psi_{\rm out}\rangle$ has matrix of coefficients
\begin{equation}
A_{\rm out}=Z^{-1/2}T^{\vphantom t}_{1}A_{\rm in}T_{2}^{\rm t},\label{Aoutdef}
\end{equation}
with normalization factor
\begin{equation}
Z={\rm Tr}\,(T_{1}^{\vphantom t}A_{\rm in}T_{2}^{\rm t})(T^{\vphantom t}_{1}A_{\rm in}T_{2}^{\rm t})^{\dagger}.\label{Zdef}
\end{equation}
(The superscript ``t'' denotes the transpose of a matrix.)

We quantify the degree of entanglement in terms of the Clauser-Horne-Shimony-Holt parameter $S$ \cite{Cla69}, which measures the maximum violation of Bell's inequality and was used in the experiment \cite{Alt02}. This parameter can be obtained from a decomposition of $|\Psi\rangle$ into a superposition of a fully entangled state (with weight $\sqrt{P}$) and a factorized state orthogonal to 
it \cite{Gisi91,Abo01}. The relation is
\begin{equation}
S=2\sqrt{1+P^{2}},\;\;P^{2}=4\,{\rm Det}A^{\vphantom{\dagger}}A^{\dagger}, \label{SPdef}
\end{equation}
with ``Det'' the determinant and $0 \le P \le 1$. (The concurrence~\cite{Hill97} is identical to $P$.) A fully  entangled state has $P=1$, $S=2\sqrt{2}$, while a factorized state has $P=0$, $S=2$. The fully entangled state could be the Bell pair $\bigl(|{\rm HV}\rangle-|{\rm VH}\rangle\bigr)/\sqrt{2}$, or any state derived from it by a local unitary transformation ($A\rightarrow UAV$ with $U,V$ arbitrary unitary matrices). 
The degree of entanglement $P_{\rm in}=2|{\rm Det}\,A_{\rm in}|$ of the incident state is given and we seek the degree of entanglement $P_{\rm out}=2|{\rm Det}\,A_{\rm out}|$ of the transmitted state. We are particularly interested in the largest $P_{\rm out}$ that can be reached by applying local unitary transformations to the incident state. This would correspond to the experimental situation that the polarizations of the two incoming photons are rotated independently, in order to maximize the violation of Bell's inequality of the detected photon pair.  

Before proceeding with the calculation we introduce some parametrizations. The Hermitian matrix product $T_{i}^{\vphantom{\dagger}}T_{i}^{\dagger}$ has the eigenvalue--eigenvector decomposition
\begin{equation}
T_{1}^{\vphantom{\dagger}}T_{1}^{\dagger}=U^{\dagger}
\! \left(\begin{array}{cc}
T_{1+}&0\\
0&T_{1-}
\end{array}\right)
\! U,\;\;
T_{2}^{\vphantom{\dagger}}T_{2}^{\dagger}=V^{\dagger}
\! \left(\begin{array}{cc}
T_{2+}&0\\
0&T_{2-}
\end{array}\right)
\! V^{\vphantom{\dagger}}.\label{decomposition}
\end{equation}
The matrices of eigenvectors $U,V$ are unitary and the transmission eigenvalues $T_{i\pm}$  are real numbers between 0 and 1. We order them such that $0 \le T_{i-} \le T_{i+} \le 1$ for each $i=1,2$. We will see that the maximal entanglement transfer depends only on the ratios $\tau_{i}={T_{i+}/T_{i-}}$. This parametrization therefore extracts the two significant real numbers $\tau_{1},\tau_{2}$ out of eight complex transmission amplitudes.  
The Hermitian matrix product $A_{\rm in}^{\vphantom{\dagger}}A_{\rm in}^{\dagger}$ has eigenvalues
$\lambda_{\pm}=\frac{1}{2}\pm\frac{1}{2}(1-P_{\rm in}^{2})^{1/2}$. These appear in the polar decomposition
\begin{equation}
UA_{\rm in}V=e^{i\phi}\!\left(
\begin{array}{cc}
u_{+}&u_{-}\\
-u_{-}^{\ast}&u_{+}^{\ast}
\end{array}\right)
\! \! \left(
\begin{array}{cc}
\sqrt{\lambda_{+}}&0\\
0&\sqrt{\lambda_{-}}
\end{array}\right)
\! \! \left(
\begin{array}{cc}
v_{+}&v_{-}\\
-v_{-}^{\ast}&v_{+}^{\ast}
\end{array}\right).
\label{parametrization}
\end{equation}
The phase $\phi$ is  real and $u_{\pm},v_{\pm}$ are complex numbers constrained by $|u_{\pm}|=(\frac{1}{2}\pm u)^{1/2}$, 
$|v_{\pm}|=(\frac{1}{2}\pm v)^{1/2}$, with real $u,v\in(-\frac{1}{2},\frac{1}{2})$. These numbers can be varied by local unitary transformations, so later on we will want to choose values which maximize the detected entanglement. 

With these parametrizations a calculation of the determinant of $A_{\rm out}$ leads to the following relation between $P_{\rm in}$ and $P_{\rm out}$:
 \begin{eqnarray}
P_{\rm out}&=&\frac{P_{\rm in}\sqrt{\tau_{1}\tau_{2}}}{(\tau_{1}-1)(\tau_{2}-1)}\bigl[\lambda_{+}Q_{+}+\lambda_{-}Q_{-}\nonumber\\
&-&2\sqrt{\lambda_{+}\lambda_{-}}({\textstyle\frac{1}{4}}-u^{2})^{1/2}({\textstyle\frac{1}{4}}-v^{2})^{1/2}\cos\Phi\bigr]^{-1},\label{Pgeneralresult}\\
Q_{\pm}&=&\left(u\pm{\textstyle\frac{1}{2}}\frac{\tau_{1}+1}{\tau_{1}-1}\right)
\left(v\pm{\textstyle\frac{1}{2}}\frac{\tau_{2}+1}{\tau_{2}-1}\right).\label{Qdef} \quad 
\end{eqnarray}
The phase $\Phi$ equals the argument of $u_{+}u_{-}^{\ast}v_{+}v_{-}$. To maximize $P_{\rm out}$ we should choose $\Phi=0$.

We first analyze this expression for the case of a fully entangled incident state, as in the experiment \cite{Alt02}. For $P_{\rm in}=1$ one has $\lambda_{+}=\lambda_{-}=1/2$, and Eq.\ (\ref{Pgeneralresult}) simplifies to
\begin{eqnarray}
P_{\rm out}&=&\frac{4\sqrt{\tau_{1}\tau_{2}}}{(\tau_{1}+1)(\tau_{2}+1)+4a(\tau_{1}-1)(\tau_{2}-1)}, \label{Presult} \\
a&=&uv-({\textstyle\frac{1}{4}}-u^{2})^{1/2}({\textstyle\frac{1}{4}}-v^{2})^{1/2}\cos\Phi. \label{a}
\end{eqnarray}
Since $\tau_{i} \ge 1$ and $|a| \le {\textstyle{\frac{1}{4}}}$ we conclude that the degree of entanglement is bounded  by $P_{\rm min} \le P_{\rm out} \le P_{\rm max}$, with
\begin{equation}
P_{\rm min}=\frac{2\sqrt{\tau_{1}\tau_{2}}}{1+\tau_{1}\tau_{2}}, \; \; 
P_{\rm max}=\frac{2\sqrt{\tau_{1}/\tau_{2}}}{1+\tau_{1}/\tau_{2}}.
\label{Pmaxdef}
\end{equation}

The maximum $P_{\rm max}$ can always be reached by a proper choice of the (fully entangled) incident state, so the maximal violation of Bell's inequality is given by
\begin{equation}
S_{\rm max}=2\sqrt{1+\frac{4\tau_{1}/\tau_{2}}{(1+\tau_{1}/\tau_{2})^{2}}}.\label{Smaxdef}
\end{equation}
The dependence of $S_{\rm max}$ on $\tau_{1}/\tau_{2}$ is plotted in Fig.\ ~\ref{smax}. Full entanglement is obtained for $\tau_{1}=\tau_{2}$, hence for
$T_{1+}T_{2-}=T_{1-}T_{2+}$. Generically,
this requires either identical objects ($T_{1\pm}=T_{2\pm}$) or non-identical objects with $T_{i+}=T_{i-}$. If $\tau_{1}=\tau_{2}$ there
are no ``which-way'' labels and entanglement fully survives with no degradation. 

Small deviations of $\tau_{1}/\tau_{2}$ from unity only reduce the entanglement to second order,
\begin{equation}
S_{\rm max}=2\sqrt{2}\bigl[ 1-\frac{1}{16}(\tau_{1}/\tau_{2}-1)^{2}+{\cal O}(\tau_{1}/\tau_{2}-1)^{3}\bigr]. \label{Smaxorder}
\end{equation}
So for a small reduction of the entanglement one can tolerate a large mismatch of the transmission 
probabilities. 
In particular, the experimental result $S=2.71$ for plasmon-assisted entanglement transfer~\cite{Alt02} can be reached with more 
than a factor two of  mismatch ($S=2.71$ for $\tau_{1}/\tau_{2}=2.4$).

As a simple example we calculate the symmetry parameter $\tau_{1}/\tau_{2}$ for a Lorentzian transmission
probability, appropriate for plasmon-assisted entanglement transfer ~\cite{Sch98,Tre99,Por99,Mar01,Rat88}.
We take
\begin{equation}
T_{i\pm}=\frac{\mathcal{T}\Gamma^{2}}{(\omega_{0}-\omega_{i\pm})^{2}+\Gamma^{2}},
\end{equation}
where $\omega_{0}$ is the frequency of the incident radiation, $\Gamma$ is
the linewidth, and $\mathcal{T}$ is the transmission probability at the resonance frequency $\omega_{i\pm}$.  
(For simplicity we take polarization-independent $\Gamma$ and $\mathcal{T}$.) 
The transmission is through an optically thick metal film with a rectangular array of subwavelength holes (lattice constants $L_{i\pm}$).
The dispersion
relation of the surface plasmons is $\omega_{i\pm}=(1+1/\epsilon)^{1/2}2\pi nc/L_{i\pm}$~\cite{Rat88},
where $\epsilon$ is the real part of the dielectric constant and $n$ is the order of the resonance, equal to the number of plasmon-field oscillations in a 
lattice constant.
We break the symmetry by taking one square array of holes and one rectangular array (lattice constants 
$L_{0}=L_{1+}=L_{2+}=L_{2-}$
and $L_{1}=L_{1-}$).
The lattice constant $L_{0}$ is chosen such that the incident radiation is at resonance. 
The symmetry parameter becomes
\begin{equation}
\frac{\tau_{1}}{\tau_{2}}=1+(2\pi)^{2}\left(\frac{nl}{L_{0}}-\frac{nl}{L_{1}}\right)^{2}, 
\quad l=\frac{c}{\Gamma}\sqrt{\frac{\epsilon +1}{\epsilon}}. \label{Sparholes}
\end{equation}
The length $l$ is the propagation length of the surface plasmon. (We have taken $c(1+1/\epsilon)^{1/2}$ for the plasmon group velocity, valid
if $\omega_{0}$ is not close to the plasma frequency ~\cite{Rat88}.)
Combining Eqs. \ (\protect\ref{Smaxorder}) and (\protect\ref{Sparholes}) we see that the deviation of $S_{\rm{max}}$ from $2\sqrt{2}$ (the degradation of
the entanglement) is proportional to the {\em{fourth}} power of the difference between the number of oscillations of the plasmon field along the
two lattice vectors.

\begin{figure}[h!]
\epsfig{file=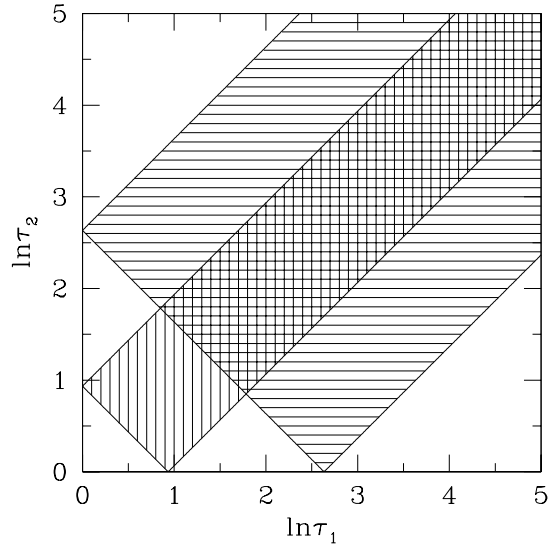,height=8cm,width=8cm}
\caption{The shaded strips indicate the values of $\ln\tau_{1}$ and $\ln\tau_{2}$ for which $P_{\rm out}=1$ can be reached with 
$P_{\rm in}=0.5$ (horizontally shaded) and $P_{\rm in}=0.9$ (vertically shaded), in accordance with Eq.\ (\protect\ref{arcos}).}
\label{regions}
\end{figure}

Turning now to the more general case of a partially entangled incident state, we ask the following question: Is it possible to achieve 
$P_{\rm out}=1$ even if $P_{\rm in}<1$? In other words, can one detect a $2\sqrt{2}$ violation of Bell's inequality after transmission 
even if 
the original state was only partially entangled? Examination of Eq.\ (\protect\ref{Pgeneralresult}) shows that the answer to this question is: 
{\em{Yes}}, provided $\tau_{1}$ and
$\tau_{2}$ satisfy
\begin{equation}
\big|\ln\frac{\tau_{1}}{\tau_{2}}\big| \le 2\,{\rm arcosh}(P_{\rm in}^{-1}) \; \mbox{and} \; 
\ln\tau_{1}\tau_{2} \ge 2\,{\rm arcosh}(P_{\rm in}^{-1}). \label{arcos} 
\end{equation}
The allowed values of $\tau_{1}$ and $\tau_{2}$ lie in a strip that is open at one end, see Fig.\ ~\ref{regions}. The boundaries are reached at 
$|u|=|v|={\textstyle{\frac{1}{2}}}$. The region inside the strip is reached by choosing both $|u|$ and $|v| < 1/2$. For $P_{\rm in}=1$ 
the
strip collapses to the single line $\tau_{1}=\tau_{2}$, in agreement with Eq.\ (\protect\ref{Pmaxdef}).

The possibility to achieve $P_{\rm out}=1$ for $P_{\rm in}<1$ is an example of distillation of entanglement~\cite{Ben96,Nie00}.
(See Ref.\ ~\cite{Lee00,Got01,Yam01,Pan01} for other schemes proposed recently, and Ref.\ ~\cite{Kwi01} for an experimental realization.) 
As it should, no entanglement is created in this operation.
Out of $N$ incoming photon-pairs with entanglement $P_{\rm in}$ 
one detects $NZ$ pairs with entanglement $P_{\rm out}=P_{\rm in}Z^{-1}\sqrt{T_{1+}T_{1-}T_{2+}T_{2-}}$, so that
$NZP_{\rm out} \le NP_{\rm in}$.

In conclusion, we have shown that optical entanglement transfer and distillation through a pair of linear media can be described
by two ratios $\tau_{1}$ and $\tau_{2}$ of polarization-dependent transmission probabilities. For fully entangled incident radiation, the maximal violation of Bell's inequality at the detectors is given by a function~(\ref{Smaxdef}) of $\tau_{1}/\tau_{2}$ which decays only slowly around the optimal value $\tau_{1}/\tau_{2}=1$. 
Distillation of a fully entangled Bell pair out of partially entangled incident radiation is possible no
matter how low the initial entanglement, provided that $\tau_{1}$ and $\tau_{2}$ satisfy the two inequalities~(\ref{arcos}).

Our results provide a simple way to describe the experiment~\cite{Alt02} on plasmon-assisted entanglement transfer, in terms of two
separately measurable parameters. By changing the square array of holes used in Ref.\ ~\cite{Alt02} into a rectangular array
(or, equivalently, by tilting the square array relative to the incident beam), one can 
move away from the point $\tau_{1}=\tau_{2}=1$ and search for the entanglement distillation predicted here. The possibility to
extract 
Bell pairs by manipulating surface plasmons may have interesting applications in quantum information processing.

This work was supported by the Dutch Science Foundation NWO/FOM and by the US Army Research Office (grant DAAD 19-02-0086). 
J.T. acknowledges the financial support provided through the European Community's Human Potential Program under contract
HPRN--CT--2000-00144, Nanoscale Dynamics.
We have benefitted from discussions with
J. Preskill and J.P. Woerdman.

\end{document}